\documentclass[%
 aip,
 amsmath,amssymb,
 reprint,%
superscriptaddress
]{revtex4-1}

\usepackage{graphicx}
\usepackage{dcolumn}
\usepackage{bm}
\usepackage{epstopdf}
\usepackage{xcolor}

\usepackage[utf8]{inputenc}
\usepackage[T1]{fontenc}
\usepackage{mathptmx, hyperref}
\usepackage{etoolbox}
\usepackage{natbib}
\usepackage{url}

\makeatletter
\def\@email#1#2{%
 \endgroup
 \patchcmd{\titleblock@produce}
  {\frontmatter@RRAPformat}
  {\frontmatter@RRAPformat{\produce@RRAP{*#1\href{mailto:#2}{#2}}}\frontmatter@RRAPformat}
  {}{}
}%
\makeatother
\begin{document}

\preprint{AIP/123-QED}

\title[Review Article]{An engineering guide to superconducting quantum circuit shielding}
\author{Elizaveta I. Malevannaya}
\affiliation{Shukhov Labs, Quantum Park, Bauman Moscow State Technical University, Moscow, 105005, Russia}
\affiliation{Dukhov Automatics Research Institute, VNIIA, Moscow, 127055, Russia}

\author{Viktor I. Polozov}
\affiliation{Shukhov Labs, Quantum Park, Bauman Moscow State Technical University, Moscow, 105005, Russia}

\author{Anton I. Ivanov}%
\affiliation{Shukhov Labs, Quantum Park, Bauman Moscow State Technical University, Moscow, 105005, Russia}
\affiliation{Dukhov Automatics Research Institute, VNIIA, Moscow, 127055, Russia}

\author{Aleksei R. Matanin}
\affiliation{Shukhov Labs, Quantum Park, Bauman Moscow State Technical University, Moscow, 105005, Russia}

\author{Nikita S. Smirnov}
\affiliation{Shukhov Labs, Quantum Park, Bauman Moscow State Technical University, Moscow, 105005, Russia}

\author{Vladimir V. Echeistov}
\affiliation{Shukhov Labs, Quantum Park, Bauman Moscow State Technical University, Moscow, 105005, Russia}

\author{Dmitry O. Moskalev}
\affiliation{Shukhov Labs, Quantum Park, Bauman Moscow State Technical University, Moscow, 105005, Russia}

\author{Dmitry A. Mikhalin}
\affiliation{Shukhov Labs, Quantum Park, Bauman Moscow State Technical University, Moscow, 105005, Russia}

\author{Denis E. Shirokov}
\affiliation{Shukhov Labs, Quantum Park, Bauman Moscow State Technical University, Moscow, 105005, Russia}

\author{Yuri V. Panfilov}
\affiliation{Shukhov Labs, Quantum Park, Bauman Moscow State Technical University, Moscow, 105005, Russia}

\author{Ilya A. Ryzhikov}
\affiliation{Shukhov Labs, Quantum Park, Bauman Moscow State Technical University, Moscow, 105005, Russia}
\affiliation{Institute of Theoretical and Applied Electrodynamics, Moscow, 125412, Russia}

\author{Aleksander V. Andriyash}
\affiliation{Dukhov Automatics Research Institute, VNIIA, Moscow, 127055, Russia}

\author{Ilya A. Rodionov}
\affiliation{Shukhov Labs, Quantum Park, Bauman Moscow State Technical University, Moscow, 105005, Russia}
\affiliation{Dukhov Automatics Research Institute, VNIIA, Moscow, 127055, Russia}
  \email{irodionov@bmstu.ru}

\date{\today}

\begin{abstract}
In this review, we provide a practical guide on protection of superconducting quantum circuits from broadband electromagnetic and infrared-radiation noise by using cryogenic shielding and filtering of microwave lines. Recently, superconducting multi-qubit processors demonstrated quantum supremacy and quantum error correction below the surface code threshold. However, the decoherence-induced loss of quantum information still remains a challenge for 100+ qubit quantum computing. Here, we review the key aspects of superconducting quantum circuits protection from stray electromagnetic fields and infrared radiation – multilayer shielding design, materials, filtering of the fridge lines and attenuation, cryogenic setup configurations, and methods for shielding efficiency evaluation developed over the last 10 years. In summary, we make recommendations for creation of an efficient and compact shielding system as well as microwave filtering for a large-scale superconducting quantum systems. 
\end{abstract}

\maketitle

\section{\label{sec:level1}INTRODUCTION}

One of the most promising physical realizations of quantum computers and simulators are superconducting quantum circuits (SQC) \cite{Nakamura1999, Blais2021, Krantz2019, Kjaergaard2020}. Nowadays, there are several superconducting quantum computer prototypes capable to solve very complex specific tasks, which are not amenable for the state-of-the-art supercomputers \cite{Arute2019, Morvan2024, Acharya2025, Gao2025, Kim2023}. However, quantum information loss or decoherence is still one of the key challenges on the way to the practically useful quantum computers. Robust error-free operation of the superconducting qubits, which are extremely sensitive to noise \cite{Bylander2011, Koch2007, Yan2012, Bialczak2007, Schuster2005}, requires minimizing both external and internal sources of noise. Internal sources can be suppressed by improving the design \cite{Ganjam2024, Pan2022}, materials \cite{Kim2021, Place2021, Oliver2013} and technology \cite{Pishchimova2023, Moskalev2023} of quantum circuits.\\
In the last two decades, the leading scientific groups in the field demonstrated tremendous results in minimizing the internal noise sources including magnetic vortices \cite{Ithier2005}, quasiparticle tunneling \cite{Córcoles2011, zaretskey2013,deVisser2011}, two-level systems \cite{Ithier2005, zaretskey2013} and dielectric losses in substrate \cite{calusine2018, wang2015,Woods2019}. A number of papers are devoted to an external noise sources suppression, such as flux and charge noise \cite{Ithier2005}, Purcell effect \cite{purcell_spontaneous_1946}, interaction with stray electromagnetic modes of a cryogenic experimental setup or environment \cite{Córcoles2011, zaretskey2013}. However, superconducting circuits protection from a broadband spontaneous radiation is currently being researched: from surrounding equipment (frequency below 300 $GHz$, wave length more 1 $mm$), infrared (IR) radiation (frequency from 300 $GHz$ to 430 $THz$, wave length from 0.74 ${\mu}m$ to 1 $mm$), cosmic rays, and background radiation (frequency above 430 $THz$, wave length less 0.74 ${\mu}m$)  \cite{Córcoles2011, Martinis2021} is currently being researched.\\
To satisfy the $kT$<<${\hbar}$${\omega}$ condition \cite{devoret2004} the superconducting qubits operating in $GHz$ frequency range, should be placed at near absolute zero temperature (in dilution refrigerators (or fridges) below 50 $mK$). Within the fridge, there are several sources of infrared radiation \cite{Krinner2019}. First, there are warmer stages of the fridge itself, and second, the passive elements of the cryogenic setup, which can dissipate electrical energy as a heat. An important factor of the decoherence is an interaction of SQC with an infrared (IR) radiation. The power transferred to SQC from IR photons is greater than from cosmic rays or background radiation  \cite{Martinis2009}. At the same time, cosmic rays and background radiation have a greater penetrating power compared to IR radiation. We briefly discuss the protection from cosmic rays and background radiation below, but this work is mostly focused on protecting against electromagnetic fields and IR radiation, which are the main sources of external sources of decoherence.\\
External radiation may destroy Cooper pairs, generating quasiparticles, which then tunnel through a Josephson junction, which in turn results in both qubit energy relaxation and dephasing \cite{Martinis2009, Catelani2011, Catelani2012}. An external magnetic field variation and induced charges near the Josephson junction area cause uncontrolled changes in the Hamiltonian parameters, specifically, the Josephson energy, $E_J$, the superconducting phase difference, ${\delta}$, and the qubit frequency, ${\omega}_{01}$ \cite{Ithier2005}. Thus, we especially focus on protecting superconducting quantum circuits from broadband external electromagnetic fields and IR-radiation.
A spontaneous radiation can either be from a free space or coaxial microwave (MW) transmission lines installed in the fridge. Depending on the path of stray radiation propagation, i.e. in a free space or by microwave lines, various shielding systems or filtering are used. The most effective shielding can be achieved by combining these two approaches. The shielding of the SQC usually represents a system of nested cylinders surrounding the sample holder with a chip. We refer the metal enclosure embedding printed circuit board with a superconducting circuit as a sample holder.\\
Articles by leading scientific groups show a wide variety of shielding systems, which differ greatly in the number of shields and their combination \cite{Place2021, Burnett2019, Kreikebaum2016, Yan2016, Grunhaupt2018, Bengtsson2020, Reagor2015, Asaad2016, Janzen2023}. Represented shielding systems comparison shows that they differ from each other by the number of shields, their relative position, position inside the dilution refrigerator, and the design and materials used.

Despite of the differences between considered shielding systems, there are also some common solutions:
\begin{enumerate}
    \item Multilayered shielding. In considered works we found a trend towards gradual increase of the shielding complexity by adding of more shielding layers. In the papers over the last 5 years, three to six shields are used in the shielding systems. However, a single-layer shielding is still frequently used, and two-layer systems comprise the majority of all shielding configurations.
    \item Shielding versatility. The considered shielding systems comprise multiple layers providing protection against both IR radiation and magnetic fields. In certain configurations, a single shield can simultaneously serve as both IR and magnetic shielding. This dual functionality is achieved by applying an IR-absorbent coating to a superconducting shield.
    \item IR shielding implementation. The IR shielding consists of a metallic substrate coated with a specialized IR-absorbing material. This coating is typically produced using epoxy resins, either pure or mixed with silicon carbide (SiC) particles or graphite powder. The primary requirement for these coatings is achieving high IR absorption efficiency.
    \item Magnetic shielding implementation. Magnetic shielding consists of superconducting shields or shields made of metals with a high level of relative magnetic permeability (typically, more than 100,000).
    \item In addition to the shields, various types of filters (including IR filters) protecting from an IR radiation are used.
\end{enumerate}
At the same time the following problems can be highlighted after review of shielding system configurations:
\begin{enumerate}
    \item There are no special guidelines for ordering the shields, recommendations for the shields’ placement inside the dilution refrigerator, instructions to choosing a number of shields.
    \item The absence of a clearly defined set of materials for the shields. Among the various materials used for shielding, the most commonly used can be identified, but there are no special instructions for their use.
    \item There is no straightforward solution to the choice of sample holder’s material and design.
\end{enumerate}
Given the wide range of possible shielding configurations and materials, it is challenging to choose an optimal shielding system, which provides the highest level of protection, while maintaining simple design and fabrication.\\
The filtering systems are also varied from one scientific group to another. The filters are an integral part of the cryogenic setup for quantum systems. They are employed in all control and measurement lines of the fridge, specifically in readout, drive, and flux lines. There are many options of the filters placement in cryogenic measurement setup (see Figure 1). In the same way as for the shielding systems, choosing the optimal filter position is also a challenging task.\\
This review represents set of requirements for choosing shielding and filtering system for effective protection of SQC from IR radiation and unwanted electromagnetic fields based on the analysis of existing multilayer shielding designs, used materials and signal lines filtering.

\begin{figure}
\includegraphics[width=0.9\linewidth]{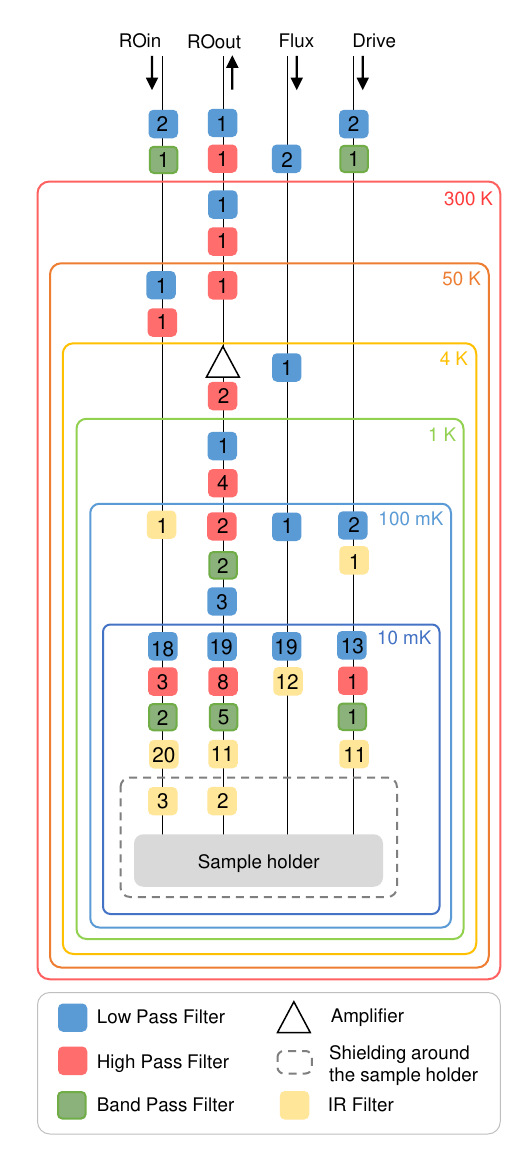}
\caption{\label{fig:epsart} Cryogenic setup with different filtering for MW and DC lines in a dilution refrigerator. The number on each element in setup represents the number of papers (from references list), which use the exact filter type at corresponding position.}
\end{figure}

\subsection{\label{sec:level2}Article organization}
In the Section II, we discuss what type of shielding is required for protection from infrared radiation and stray electromagnetic fields. We present the requirements for an effective shielding system based on theoretical estimations and numerical modeling, and analyze existing shielding systems according to these requirements.\\
In Section III, we discuss microwave lines filtering. Specifically, we describe the requirements for the filters used in cryogenic setup for quantum systems. We explore the types of filters employed in measurement circuits and their placement in different measurement lines.\\
In Section IV, we discuss the methods for evaluating the effectiveness of shielding systems and compare different shields.\\
In summary and conclusions we consolidate our recommendations for fabricating compact and efficient shielding systems to protect quantum circuits from external electromagnetic fields.

\section{SHIELDING}
Generally, the shielding system solves two main problems: absorption of an incident IR radiation and protection from stray magnetic fields. In order to design an effective shielding system, it is essential to consider these problems individually: starting from shielding principles to the materials and designs used.
\subsection{\label{sec:level2}Protection from IR radiation}
\subsubsection{\label{sec:level3}Shielding principle}
Protection from an IR radiation is based on the absorption of the energy from incident photons by the absorbing material of a shield and the conversion of this energy into internal thermal energy of the shield. This thermal energy is then removed by the cooling mechanisms within the fridge. As a result, the photons partially or completely lose the energy and can’t cause Cooper pairs braking in superconductor. To protect from the described source of an IR radiation, the shields and, in some cases, sample holder lids are coated by special IR absorbing materials. Commercially available epoxy resins, such as Stycast 2850 FT and Eccosorb CR series, are used as IR-absorbing coatings, either alone or in combination with other particles such as silicon carbide powder, carbon powder, or graphite dust.
\subsubsection{\label{sec:level3}Requirements for IR shielding}
Guided by radiative heat transfer theory, we derive the core requirements for effective IR shielding of quantum devices:

\begin{enumerate}
  \item Optimal IR radiation shielding requires configuring the shield's inner surface with an absorption coefficient ($A$) approaching 1 to effectively absorb penetrating radiation, while the outer surface should maintain $A {\approx} 0$ to maximize reflection of incident external radiation.
  \item To maximize reflection of incident IR radiation, the shield's outer surface should be fabricated from a polished, highly reflective metal.
  \item The absorbing coating on the inner shield side should have an extended effective absorbing surface and high thermal conductivity for rapid heat dissipation.
  \item To increase the absorption coatings capacity one should insure its surface roughness on the order of the IR wavelength \cite{Smith1984}. This level of roughness avoid specular reflection, scattering the incident IR radiation, which undergoes multiple reflections and corresponding interactions within absorption material, extending an effective surface and overall absorption. In cryogenic quantum applications a far-IR range with wavelengths of 700 – 1000 ${\mu}$$m$ become one of the most important \cite{Ekin_cryostat}, consequently, macroscopic roughness of the same size should be introduced for better absorption capacity.
  \item The temperature of the shield must be lower than the temperature of the quantum circuit to ensure the heat transfer from the sample to the shield. The shield base should be made of a metal with the highest possible thermal conductivity at operating temperature and have robust thermal contact between the shield and the lower stage of the fridge.
\end{enumerate}

Following these requirements is the proper way for creating an effective IR shielding. Let’s consider the existing IR shields designs.
\subsubsection{\label{sec:level3}Practical implementation of IR shielding}
\paragraph{Material for the IR shields base.}
The requirements for IR shielding shows, that it is necessary to use a shield base made of material with a high thermal conductivity. The thermal conductivity of some materials commonly used for the base of IR shields is represented in Table 1 below.\\
As shown in Table 1, high purity copper has a higher thermal conductivity, making it preferable material for the base of IR shielding.\\
The use of copper for the shielding base is also supported by shielding systems described in the literature. Figure 2 shows the most commonly used materials for the base of IR shielding described in \cite{Schwartz2016, Chou2018, Dickel2018, Monroe2021}. Approximately 65\% \footnote{Here and further, the estimate of mention frequency of a certain parameter is given from the total number of works where this parameter is found} of shielding configurations (Figure 2) integrate combined IR/magnetic shielding, using either copper substrates with superconducting coatings (aluminum (Al) \cite{Grunhaupt2018, henriques2019}, indium (In) \cite{Premkumar2021}, or tin (Sn) \cite{Kreikebaum2016}) or bulk superconductors \cite{Smith2019, Zhang2018,Gyenis2021, Hazard2019}. This combined shielding approach enables compact and lightweight shielding designs. However, pure superconducting shields exhibit significantly lower heat dissipation capacity than their copper-based counterparts.\\

\begin{table}
\caption{\label{tab:table4}Thermal conductivity of materials used for IR shielding.}
\begin{ruledtabular}
\begin{tabular}{lcr}
Material&\mbox{${\kappa}$, at 300 $K$,}&\mbox{${\kappa}$, at 4 $K$}\\
\textcolor{white}{second line}&\mbox{$W\cdot m^{-1}\cdot K^{-1}$}&\mbox{$W\cdot m^{-1}\cdot K^{-1}$}\\
\hline
Copper (99.999\%)&\mbox{${\sim}$400 \cite{Ekin_cryostat}}&\mbox{${\sim}$15,000 \cite{Ekin_cryostat}}\\
Gold&\mbox{310 \cite{AZO_materials}}&\mbox{-}\\
Aluminum (99.999\%) & \mbox{${\sim}$400 \cite{Ekin_cryostat}} & \mbox{${\sim}$3,000 \cite{Ekin_cryostat}}\\
\end{tabular}
\end{ruledtabular}
\end{table}

Rarely, gold-plated copper \cite{ByoungMoo2021, Wu2021} is used as the base material for the IR shield. Gold-plating of the entire shielding components does not provide any advantages, as it reduces thermal conductivity at cryogenic temperatures compared to uncoated copper. However, gold-plating is highly beneficial for thermal interfaces between shielding components. Gold-gold contacts demonstrate over an order of magnitude greater thermal conductance than copper-copper contacts \cite{Ekin_cryostat} due to the absence of surface oxidation and improved conformability \cite{Dhuley2019}. This enhancement relaxes the required surface finishing quality and contact pressure for copper interfaces.

\begin{figure}
\includegraphics[width=0.95\linewidth]{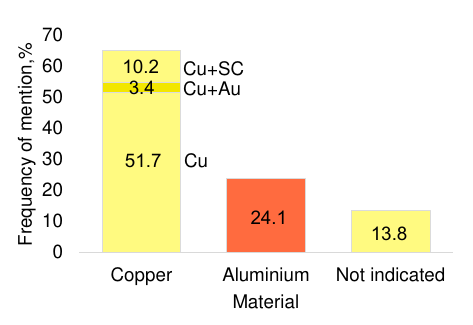}
\caption{\label{fig:epsart} Typical materials for the base of IR shields.}
\end{figure}

\paragraph{Absorbing coating for IR shielding.}
Another important component of IR shielding is an IR absorbing coating. In reference \cite{Bründermann2012}, some absorbing coatings data, that are used in the IR radiation range, are provided (Eccosorb CR110, Herberts 1356H, TK RAM, and SiC grains in Stycast). Assuming that the IR shielding material does not transmit radiation, we can consider the sum of absorption and reflection coefficients to be 1. This allows us to estimate the absorption properties based on the reflection coefficient. According to theory \cite{howell2010thermal}, absorbent coatings should exhibit maximum absorption (near 1) in the operating range. Among the absorbent coatings shown in reference \cite{Bründermann2012}, SiC grains in Stycast have the highest absorption capacity. The literature demonstrates the prevalence of Stycast epoxy-based coatings for IR shielding. These coatings are employed both in pure form \cite{Grunhaupt2018, Maleeva2018} and when modified with additives to enhance surface roughness \cite{barends2011} (Figure 3). For instance, mixtures of Stycast epoxy with carbon lampblack and 350 ${\mu}m$ SiC grit demonstrate absorption coefficients of 0.9 in the $THz$ frequency range. \cite{barends2011}. To create an optimal absorbing surface texture, various combinations \cite{Burnett2019, ByoungMoo2021} of SiC particles of different sizes \cite{Dickel2018, Jin2015, Hazard2019} and carbon powder \cite{Chou2018, Wang2022} may be utilized. Among these, the 1 $mm$ SiC particles dispersed in Stycast epoxy are particularly advantageous since their superior absorption performance. Such a specific particle size meets the surface roughness requirements for optimal IR absorption. Alternative absorber materials, such as Eccosorb resin epoxy \cite{Zhang2018, Mundada2021}, may also be considered for these applications.

\begin{figure}
\includegraphics[width=0.95\linewidth]{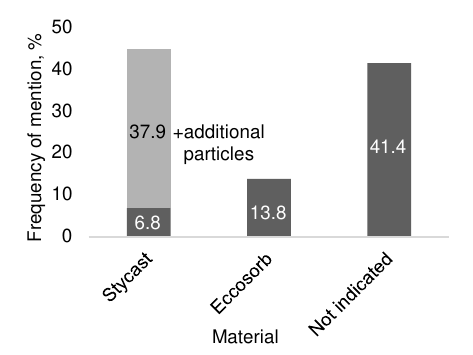}
\caption{\label{fig:epsart} Absorptive layer materials for an IR shielding.}
\end{figure}

\paragraph{Reflection layers.}
In addition to IR protection, thin metallic layers are used to reflect spontaneous parasitic IR radiation. For this application, aluminized polyethylene terephthalate (commercially known as Mylar®) is typically employed \cite{Place2021, Gyenis2021, Narla2017}. The placement of this layer may vary: it may be placed around the holder  \cite{Place2021}, around the first shield \cite{Hazard2019}, between the shields \cite{Premkumar2021, Gyenis2021, Hazard2019} or surround the entire shielding system \cite{Place2021, Premkumar2021, Hazard2019}. At the same time, the outer surface of the shield may act as a reflective layer due to its low absorption coefficient. This allows for the creation of a more compact shielding system.
\paragraph{The number of layers, the shape, and the arrangement of the IR shields.}
A more effective shielding system can be defined using the theory of a heat transfer by radiation and modeling the radiation propagation in the "sample-shielding-dilution refrigerator" system. A calculation scheme is represented in Figure 4a. In this system, "Body 1" is the sample, "Body 2" is the dilution refrigerator or radiation source, and "Shields" is the shielding system, which effectiveness we’d like to estimate. The power of a heat transfer by radiation is calculated by the following relation:

\begin{equation}
 Q_{(1,2)sh} = {\sigma}A_{(1,2)sh}F_1(T_1^4 - T_2^4),
\end{equation}

where ${\sigma}$ is the Stefan-Boltzmann constant equal to $5.67\cdot10^{-8}$ $W\cdot m^{-2}\cdot K^{-4}$; $F_1$ is the surface area of Body 1; $T_1$ and $T_2$ are the temperatures of Bodies 1 and 2, respectively; and $A_{(1,2)sh}$ is the reduced absorption coefficient of the system (taking into account the shields):
$A_{(1,2)sh} = \frac{1}{\frac{1}{A_{1,2}}+\sum\limits_{i=1}^n \frac{F_1}{F_{shi}}\left( {\frac{2}{A_{shi}}-1}\right)};$

where $F_{sh i}$ is surface area of i-th shield; $A_{sh i}$ is absorption coefficient of i-th shield; $A_{1,2}$ is reduced absorption coefficient between Body 1 and Body 2:
$A_{1,2} = \frac{1}{\frac{1}{A_1}+\frac{F_1}{F_2}\left( {\frac{1}{A_1}-1}\right)};$

where $A_1$ and $A_2$ are the absorption coefficients of Body 1 and Body 2, respectively; and $F_2$ is the surface area of Body 2.

In case the shield has different absorption coefficients of its inner and outer surfaces, formula for $A_{(1,2)sh}$ will be the following: $A_{(1,2)sh} = \frac{1}{\frac{1}{A_{1,2}}+\sum\limits_{i=1}^n \frac{F_1}{F_{shi}}\left( {\frac{1}{A_{shi(inner)}}+\frac{1}{A_{shi(outer)}}-1}\right)};$

where $A_{shi(inner)}$ and $A_{shi(outer)}$ - the absorption coefficients of inner and outer surfaces of the i-th shield, respectively.

Using these formulas, it is possible to determine the IR shielding configuration, that provides the minimum temperature of Body 1 ($T_1$), i.e. a sample with a quantum circuit.

\begin{figure}
\includegraphics[width=0.8\linewidth]{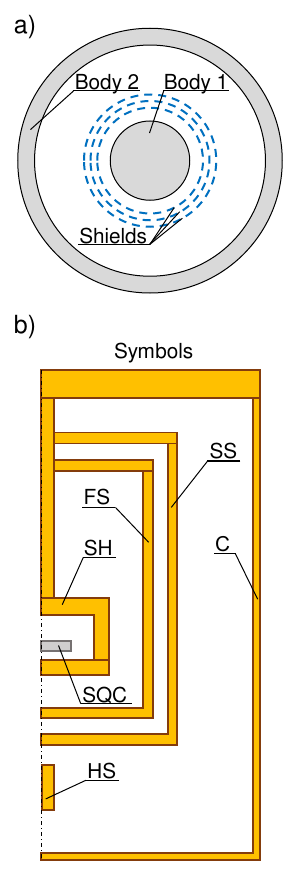}
\caption{\label{fig:epsart} a) Schematic of the radiative heat transfer model between two bodies with multiple shielding elements. b) Two-dimensional axisymmetric representation of the shielding system simulation. Key components: SQC (superconducting quantum circuit), SH (sample holder), FS (first shield), SS (second shield), C (cryostat's lower-stage can), and HS (heat source).}
\end{figure}

\begin{figure*}
\includegraphics{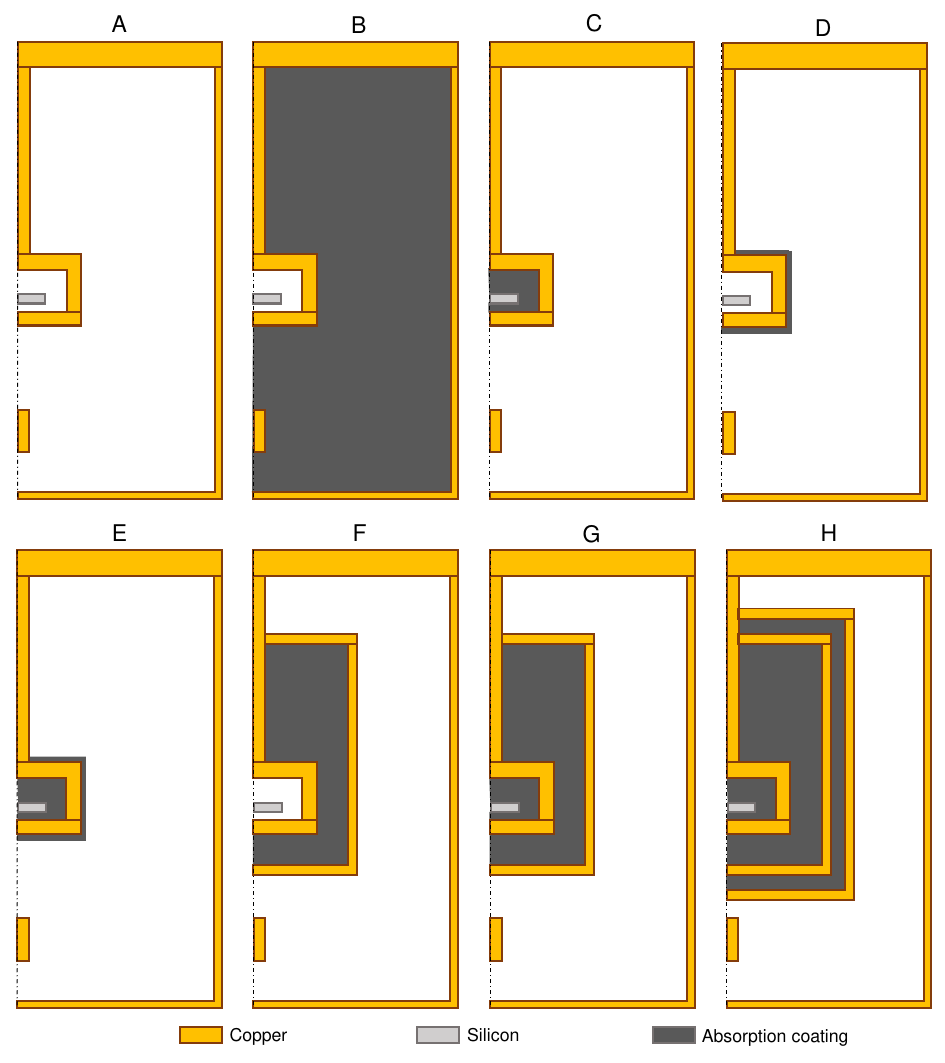}
\caption{\label{fig:wide}Schematic of the analyzed shielding configurations: A – uncoated copper sample holder with copper can attached to the lowest stage of a fridge; B –sample holder with absorption-coated can; C –sample holder with an inner absorption coating and uncoated can; D – sample holder with  outer absorption coating and uncoated can; E – sample holder with both inner and outer coating and  uncoated can; F – uncoated sample holder with inner coated first shield; G – inner coated sample holder with inner coated  first shield and uncoated can; H – inner coated sample holder with two inner coated shields and uncoated can.}
\end{figure*}

The results of the calculations is verified using finite element simulations. Computer calculations is performed using COMSOL Multiphysics Heat Transfer in Solids and Surface to Surface Radiation. A two-dimensional axisymmetric model (see Figure 4b) was constructed to simplify calculations.In Figure 4b we introduce the following designations: “SQC” – superconducting quantum circuit, “SH” – sample holder, “FS” – first shield, “SS” – second shield, “HS” – heat source in the form of an IR filter, and “C” – can on the lowest stage of the fridge.\\
In our model, we utilize identical surface absorption coefficients for all computational scenarios. The absorption coefficient for silicon ($A_{Si}$) is 0.00008 \cite{Wollack2020}, while for aluminum ($A_{Al}$) it is 0.0012 \cite{Benbow1975}. The copper coefficient ($A_{Cu}$) is set to 0.005 \cite{Biondi1956}, and the absorptive coating ($A_{Absorp.coat.}$) has a value of 0.9 \cite{barends2011}. The system operates with a fixed heat dissipation power of $1\cdot10^{-14} W$ (this evaluation has been done on the base of heat loads calculations carried out in Ref.\cite{Krinner2019}). The temperature of the cryostat's lower stage is maintained at 0.01 $K$ throughout the simulations. We examine two distinct thermal scenarios: one where the superconducting quantum circuit itself serves as the heat source (designated SQC in Figure 4b), and another where an external heat source (designated HS in Figure 4b) provides the thermal input.\\
We considered eight shielding configurations ordered by complexity increase (weight and cost) (Figure 5). Shielding configuration "A" consists solely of the sample holder and the can on the lower stage. This represents the simplest possible configuration. Shielding system "B" incorporates an absorbing coating on the can. Comparison with system "A" enables evaluation of thermal radiation absorption effectiveness of the lower cryostat stage shield. Configurations "C"-"E" examine different absorbing coating placements: "C" – Absorbing coating on the inner surface of sample holder; "D" - outer surface of sample holder; "E" - both surfaces. These configurations, compared with "A", determine optimal coating placement on the sample holder. Configuration "F" is the most frequently reported variant of IR shielding and it consists of the sample holder without absorbing coating surrounded by a shielded enclosure with an inner absorbing coating. Configuration "G" incorporates the same inner-coated enclosure, but applied to the optimal sample holder design (selected from configurations "A" and "C"-"E"). This arrangement allows evaluation of the effectiveness of the shield with absorbing coating surrounding the holder and comparison with widespread shielding system "F". Configuration "H" introduces a second shielded layer with absorbing coating. Configuration "H" in comparison with "G" assesses the benefit of additional shielding layers.\\
In both thermal scenarios sample temperature in the steady-state regime was evaluated using theoretical calculation and numerical modelling (Figure 6). The shielding system efficiency can be estimated by the minimum attainable temperature of the sample. In order to consider only radiative heat transfer the sample was deliberately decoupled from the refrigerator lowest plate by omitting thermal conduction paths. As a result, in each scenario sample temperature was greater than 260 $mK$ that significantly exceeds 10 $mK$ on the refrigirator's lowest stage and is not usually observed in experiments. However, this approach provides a sensitive method for reviewing the performance of shielding configurations.\\
\begin{figure}
\includegraphics[width=0.95\linewidth]{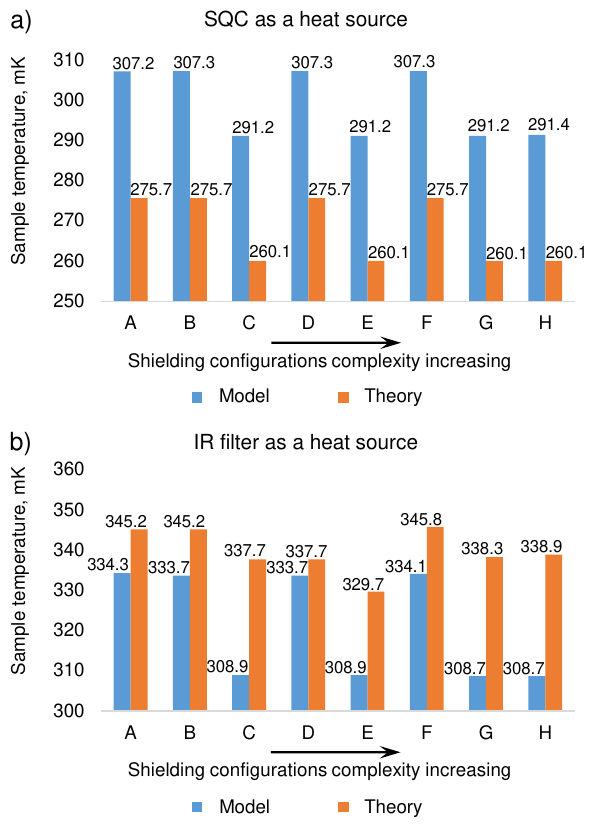}
\caption{\label{fig:epsart} Steady-state sample temperature comparison across shielding configurations: a) Internal heating case with the superconducting quantum circuit itself as the heat source; b) External heating case with an absorptive low-pass filter positioned near the sample mount as the heat source.}
\end{figure}
Our results demonstrate that the sample temperature is most strongly influenced by the nearest surface – the sample holder lid. This conclusion establishes the sample holder itself as a critical component of the shielding system that must be optimized alongside traditional radiation shields. Inner coating of the sample mount should be considered optimal. It provides efficient heat dissipation path for SQC which is crucial for "floating" qubits where the qubit capacitor pad is electrically isolated. Outer coating on a single absorptive-coated shield around the mount is also optimal. It ensures effective thermal management of the parasitic heat sources, e.g. control wiring and filters. However, the simulation reveals a performance plateau starting from the "E" configuration, where additional shielding layers (configurations "G", and "H") do not provide any improvements. While basic infrared shielding is essential, multilayer shielding leads to unnecessary design complexity that increases system mass and footprint without improving thermal isolation.\\ 
Therefore, the following recommendations for shielding configurations are proposed:
\begin{enumerate}
    \item Floating architecture qubits with nearby IR filters:\\
Configuration “G” (inner-coated sample holder + single absorptive shield) provides optimal thermal isolation from both intrinsic qubit heating and nearby filter radiation.
    \item Floating architecture without IR filters in a proximity:\\
Configuration “C” (inner-coated sample holder) offers sufficient protection while minimizing system complexity.
    \item Grounded qubit architectures with adjacent passive components:\\
Configuration “F” (uncoated holder + single absorptive shield) effectively protects from parasitic heat loads from control wiring and other components.
\item Minimal requirements (no floating elements or warm components):\\
Configuration “A” (basic uncoated copper holder) provides adequate baseline shielding without unnecessary mass or complexity.
\end{enumerate}

Optimal SQC shielding requires careful geometry selection alongside absorber placement. Our analysis shows that about 35\% of designs use light-tight\cite{Hazard2019} cup-lid configurations\cite{Place2021, Zhang2018, Grunhaupt2018, Reagor2015}, often with indium-sealed or joints\cite{Harrington2020}. Cylindrical inserts are less common \cite{ByoungMoo2021, Chou2018, Serniak2019, Wang2019}. Cup-lid configuration can be easily integrated in standard dilution refrigerators. Hermetic RF and DC connectors can be used to allow control wiring inside such a shielding. However, a vacuum pumping path with IR-tight design should be introduce to ensure efficient evacuation.\\
\paragraph{Sample holder as an essential component of IR shielding.}
We have previously demonstrated that sample holders play an important role in IR shielding. Therefore, let’s consider various types of sample holders that are described in the literature.\\
In order to ensure a robust thermal contact, approximately in a half of considered papers, oxygen-free copper \cite{Buks2020, Eddins2019, Harrington2020} is used (see Figure 7). The choice of a copper is justified by its high thermal conductivity. The holders manufactured of superconductors, such as aluminum \cite{Jin2015, Burkhart2021, Ribeill2016, Kelly2015}, considered in the other half of papers. The use of superconductors creates additional magnetic shielding owing to the Meissner effect (more detail consideration of this effect will be done later). Despite the added protection from the magnetic field, superconducting sample holders have a lower thermal conductivity compared to the copper sample holders, that is why the use of the superconductors is undesirable.  There are also copper sample holders coated by superconductor (Al \cite{Place2021} or In\cite{Narla2017}). These sample holders create additional magnetic shielding while maintaining the high thermal conductivity. However, these examples are not common. There are also a few cases, where gold-coated copper is used \cite{Lienhard2019, Blok2021}. The gold coating prevents the formation of the oxides on the copper, thereby reducing dielectric losses \cite{Huang2021}.

\begin{figure}
\includegraphics[width=0.95\linewidth]{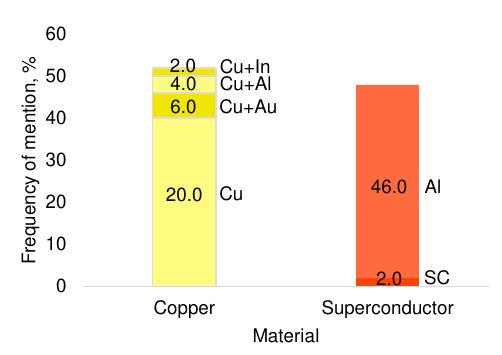}
\caption{\label{fig:epsart} Typical materials used for sample holders.}
\end{figure}

Among studies documenting sample holder geometry, more than 50\% feature transmon qubits mounted within three-dimensional cavities \cite{monroe2021optical, Serniak2019, Smith2019, Grunhaupt2018, Martin2020, Chapman2023} that simultaneously function as both chip holders and readout circuit elements. The remaining implementations employ three distinct holder configurations: approximately 9\% utilize rectangular holders\cite{Maleeva2018, Ribeill2016}, 3\% incorporate cylindrical holders\cite{Chen2018, Quintana2017}, and 27\% employ flat copper bases with printed circuit boards and protective lids positioned above the qubit chip.\cite{ByoungMoo2021, Arute2019, Zhang2018}.\\
The selection of a sample holder's shape depends on the SQC design and the wiring method to the cryogenic setup. Only the quality (roughness and absorption coefficient) of the sample holder internal surface and its sealing are of importance.\\
In 8\% of the surveyed studies, the sample holder was hermetically sealed using indium wire to achieve light-tight conditions\cite{Chou2018, Schwartz2016, Narla2017}, while only 1.6\% incorporated lids with IR-absorbent coatings \cite{Kreikebaum2016}. For the latter case, the coating consisted of Stycast epoxy mixed with carbon lampblack and 350 ${\mu}m$ SiC particles. These experimental implementations show good agreement with our prior theoretical calculations.\\
The absorbing coating on the inner surface of the sample holder allows for more efficient cooling of the SQC, but only when the holder itself has high thermal conductivity. However, high dielectric loss of the absorber can affect qubit`s coherence if they are close to each other. IR absorber should be positioned sufficiently far from the qubit. Sufficient distance can be determined using surface participation method \cite{Lienhard2019, Smirnov2024}. This approach shows that the minimal qubit-absorber distance should exceed 4 $mm$. The surface participation method allows to quantify the extent to which the absorber interacts with the qubit's field in terms of maximal possible coherence time $T_1$ that can be achieved in presence of the dielectric loss channel:
\begin{equation}
\frac{1}{T_1}=\frac{\omega}{Q}={\omega}p_{abs}tan{\delta}_{abs}+{\Gamma}_0;
\end{equation}
where $T_1$, $\omega$, and $Q$ are coherence time, angular frequency and quality factor, respectively. $tan{\delta}_{abs}$ is the dielectric loss tangent of the absorber, $p_{abs}$ is a participation ratio of the material (a fraction of electromagnetic field energy stored within this material). ${\Gamma}_0$ represents other loss channels. The simulation conducted with a typically sized floating transmon \cite{Smirnov2024} and a usual transmon on a lossless substrate in a lossless chip cavity to ensure ${\Gamma}_0=0$. Inner side of the top lid was covered with RF absorber with thickness of 0.5 $mm$ and $tan{\delta}_{abs}$ of 0.05, which is an upper bound for Stycast IR absorber. During the simulation distance between the qubit and the bottom surface of the absorber was varied from 0.5 to 5 $mm$. The modelling shows that the influence of the absorber becomes as low as an influence of dielectric losses in silicon (limiting $T_1$ more than 10 $ms$) substrate when Stycast is more than 4 $mm$ away from the qubit (see Figure 8). This ensures that the benefits of efficient cooling provided by the absorber are not compromised by degradation in qubit coherence.

\begin{figure*}
\includegraphics[width=0.95\linewidth]{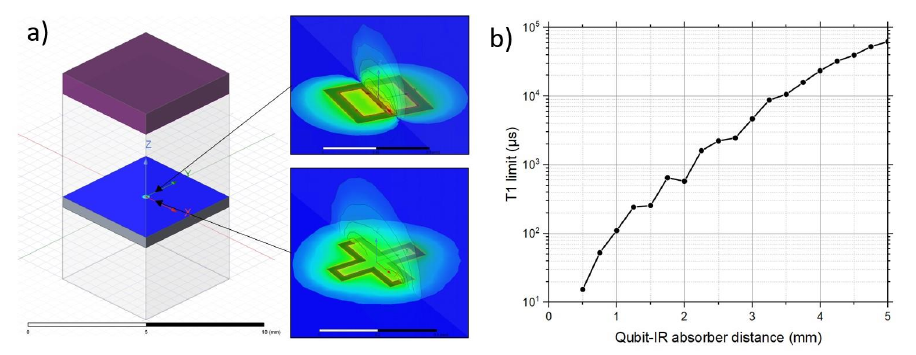}
\caption{\label{fig:wide} a) Electric field distribution in superconducting qubit cavity for transmons of various types. Length of the size mark on transmon pictures are 0.8 $mm$ for floating qubit (upper inset) and 0.4 $mm$ for transmon qubit (lower inset). IR absorber is depicted as violet area; b) Maximal possible $T_1$ time vs qubit-IR absorber distance.}
\end{figure*}

Based on comprehensive radiation modeling and literature review, we establish three critical design requirements for optimal sample holder performance:
\begin{enumerate}
    \item Copper (or copper with superconducting coating) provides optimal thermal conductivity for efficient SQC cooling, balancing both heat dissipation and magnetic shielding requirements.
    \item Indium wire gaskets offer reliable hermetic sealing of the holder lid, preventing photon leakage while maintaining cryogenic compatibility down to millikelvin temperatures.
    \item The inner surfaces of holders for floating-architecture SQCs require absorptive coatings to mitigate infrared radiation while preserving quantum coherence.
\end{enumerate}

\begin{figure*}
\includegraphics[width=0.95\linewidth]{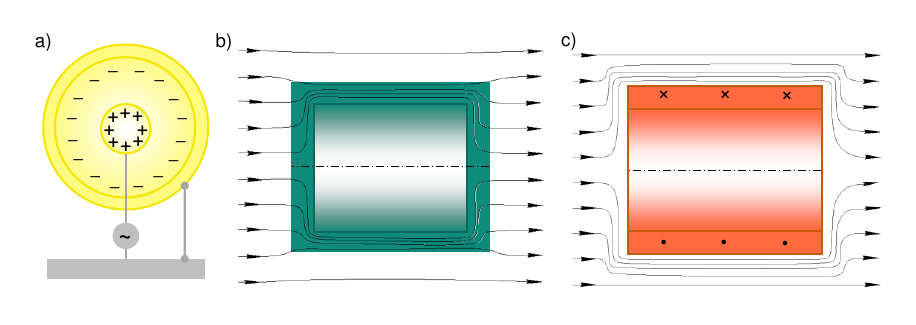}
\caption{\label{fig:wide} The shields operating on different physical principles: a) electrostatic – the external shield is discharged due to the connection to the ground; b) magnetostatic – magnetic field is concentrated in the ferromagnetic loop; c) electromagnetic –external alternating electromagnetic field, induces eddy currents in the shield loop which compensate electromagnetic field inside the shield.}
\end{figure*}

\subsection{\label{sec:level2}Protection from electrical and magnetic fields}
\subsubsection{\label{sec:level3}Shielding principle}
Generally, protection from electrical and magnetic fields consists of three components: electrostatic shielding, magnetostatic shielding, and electromagnetic shielding (Figure 9). The principle of shielding from an electrical field (Figure 9a) involves the accumulation of the charge from a free space on the shield and its subsequent leakage into the ground. For effective protection against an electrical field, the shield must be made of a highly conductive material, such as copper or aluminum, with high-quality electrical grounding to minimize the contact resistance with the ground.\\
The shields made of ferromagnetic materials, such as permalloys (${\mu}$-metal, a soft magnetic alloy, or steels), with a high relative magnetic permeability (${\mu}_r$) are used to protect from a static and slowly varying (up to 1 $kHz$) magnetic field. In these shields, the lines of magnetic flux pass along the walls (see Figure 9b), which have a lower magnetic resistance than the surrounding air. The shielding effectiveness is determined by magnetic permeability of the material the shield is made of.\\
The operation principle of the shields protecting from an alternating high-frequency magnetic field is based on an alternating electromagnetic field excitation within the shield. It creates alternating induction eddy currents (also known as Foucault currents). These currents shield the volume which this currents surround: the field inside the shield is brought down, while the field outside the shield is enhanced (see Figure 9c). As a result, the field strength is reduced inside the shield and increased outside, thus leading to the field to be pushed from the shield. This type of shielding depends on the depth of field penetration at various frequencies (known as the "skin depth"), and becomes effective above frequencies ($f$) 1 $kHz$. The lower the skin depth ($\delta$) the thinner (the lighter) the shield can be made:
\begin{equation}
    {\delta}=\sqrt{\frac{\rho}{\pi f \mu}}.
\end{equation}
Non-magnetic materials with low resistivity ($\rho$) or ferromagnetic properties with high permeability ($\mu$), such as copper, aluminum, $\mu$-metal, and steel are used to design the shields, which resist the penetration of the EM field. For example, skin depth of the copper at 100 $kHz$ equals ~200 $\mu m$.\\
Superconductors also provide excellent protection against alternating electromagnetic fields, as in an ideal conductor the currents flow on the surface without penetrating deeply into the metal, i.e. low skin depth. Ideally, such shields should have no holes or slots.\\

\subsubsection{\label{sec:level3}Requirements for shielding from electromagnetic fields}
Based on the discussed above theory of protection from electromagnetic fields, the following requirements can be summarized:

\begin{enumerate}
\item The shielding should protect from both electric and magnetic fields including constant and alternating high-frequency fields.
\item For electrostatic shielding, it is recommended to use the materials with a high electrical conductivity and robust grounding properties. Recommended materials are copper and aluminum, as they have high thermal conductivity, provide RF shielding and are widely available.
\item For magnetic shielding, it is recommended to use a metal with a high relative magnetic permeability (greater than or equal 100,000).
\item For electromagnetic shielding, it is recommended to use a superconductor (such as aluminum) or a non-magnetic metal with a low skin depth (for example, copper or aluminum).
\end{enumerate}

\subsubsection{\label{sec:level3}Practical implementation of electromagnetic shielding}
Now, we will discuss various implementations of electromagnetic shielding that are currently known.
\paragraph{Superconducting shields.}
Superconducting shields used for protection from parasitic magnetic fields are primarily manufactured from aluminum (Figure 10) \cite{Jin2015, moghaddam2019, Asaad2016, Liu2016PhD, Singh2025, Warren2023, Norris2024}. Additional superconducting materials identified in the literature include tin (Sn)\cite{Kreikebaum2016} and indium (In)\cite{Premkumar2021}. Aluminum is preferred material for superconducting shielding due to the ease of its workability and relatively low cost compared to other materials.\\
In considered papers, almost a quarter of all superconducting shields in combination with an IR shielding, which are coated with an IR-absorbing material, are used \cite{Zhang2018, Arute2019, Kreikebaum2016}. These shields usually consist of a light tight cup with a lid \cite{Place2021, Zhang2018}, which allows for more compact shielding due to the possibility of combining together IR and magnetic shielding (shielding from magnetic field). However, superconductors have a lower thermal conductivity than in its normal state. It has a negative impact on the absorbed heat dissipation, which is crucial for IR shielding. To effectively cool a superconducting shield, a superconducting layer can be placed on a copper base \cite{Premkumar2021, Kreikebaum2016}.

\begin{figure}
\includegraphics[width=0.95\linewidth]{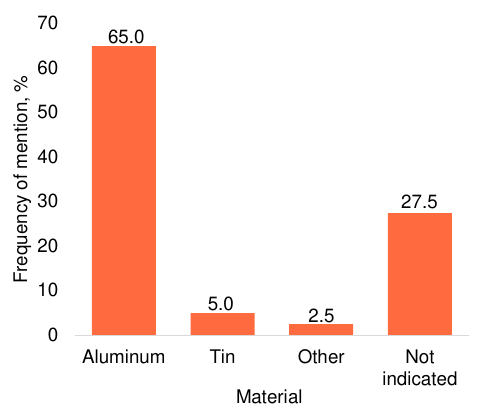}
\caption{\label{fig:epsart} Typical materials used for superconducting shields.}
\end{figure}

Cup-shaped shields without a lid \cite{Burnett2019, Bengtsson2020} or individual shields placed over the chip \cite{Arute2019} have also been presented. However, for a more uniform distribution of the magnetic field inside the shield, it is necessary to use tightly closed shields.\\
Based on the above analysis, we can conclude that a superconducting shield is an essential part for the shielding system. This shield can be realized by two methods:\\

\begin{enumerate}
\item A light-tight, one-layered aluminum superconducting shield surrounding the sample holder. This shield must follow after an IR shield.\\
\item A combined shield protecting from IR radiation and electromagnetic field, consisting of a superconducting layer on a copper light-tight shield coated by an absorbing material and surrounding the sample holder.
\end{enumerate}

\paragraph{Material for magnetic shields.}
The operation principle of magnetic shielding is based on the magnetic field retraction into a shield material that has a high relative magnetic permeability (typically 100,000 or more) \cite{Huang2021}. These components are typically fabricated from permalloy (iron-nickel alloys) that undergo specialized processing, including hydrogen atmosphere annealing. In the literature, these materials are often referred to as ${\mu}$-metals \cite{Hazard2019, Leonard2018, Andersen2019, Koolstra2022, Wilen2021} or they may have specific names (Figure 11), such as Cryoperm \cite{Lisenfeld2019, Sung2018, Ding2023, Dodge2023}, Cryophy \cite{Dickel2018, Asaad2016, Norris2024}, Amumetal \cite{Wang2022, Narla2017, Liu2016PhD, Bejanin2016}.\\
The presented materials differ in the method of their thermal processing in a hydrogen environment, and consequently, in their relative magnetic permeability (Table 2). For these shields, it is preferable to use a material with the highest relative magnetic permeability.\\
For magnetic shielding applications, cup-shaped shield configurations are commonly used \cite{Bengtsson2020, Norris2024}. Similarly, cylindrical shields with lids demonstrate comparable prevalence in the literature \cite{Bejanin2016, Quintana2017, Zhang2018}. Individual magnetic shielding is used in special cases \cite{Arute2019} As with superconducting shielding, the presence of a lid on the shield enables the equalization of the magnetic field within the protected space, thus it is recommended to use the magnetic shield with a lid.

\begin{figure}
\includegraphics[width=0.95\linewidth]{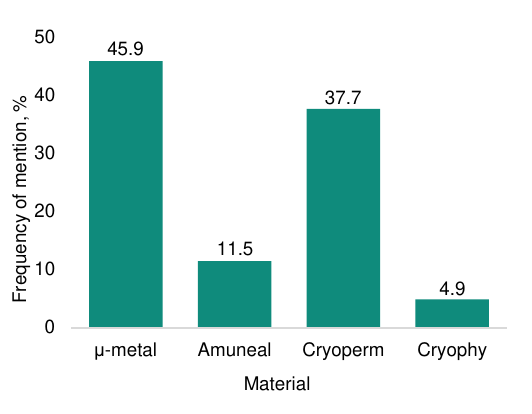}
\caption{\label{fig:epsart} Typical materials used for magnetic shields.}
\end{figure}

\begin{table}
\caption{\label{tab:table4}The relative magnetic permeability of various magnetic materials. }
\begin{ruledtabular}
\begin{tabular}{lr}
Material&\mbox{Relative magnetic permeability, ${\mu}_r$}\\
\hline
${\mu}$-metal&\mbox{80,000-400,000 \cite{Mu_metals}}\\
Cryoperm 10&\mbox{65,000-250,000 \cite{Amuneal}}\\
Amumetal& \mbox{60,000-400,000 \cite{Amuneal}}\\
Cryophy& \mbox{- \cite{Cryophy}}\\
\end{tabular}
\end{ruledtabular}
\end{table}

In some studies, specific guidelines are provided for the use of magnetic shields. For example, it is recommended not to include magnetic components inside the shield \cite{Janzen2023, Reagor2015, Opremcak2021, Gao2018}, or the sample holder should be placed closer to the bottom of the shield \cite{Underwood2015}, and a thermal anchor should be installed on the shield to improve its cooling \cite{Zhang2018, Reagor2015, Place2021}. Some magnetic shields have special design features such as waveguides on the lid \cite{ByoungMoo2021} or light tightness \cite{Hazard2019}. These design features help to minimize the influence of the magnetic field on the SQC. Almost a half of considered shielding systems have the sample holders with special features. Either the holder itself is made of a superconductor \cite{Chen2018, Mundhada2020, Wilen2021}, or it has superconducting \cite{Place2021} or absorbing layers \cite{Kreikebaum2016}. These aspects have been discussed in more details before in a section about sample holders.

\begin{figure*}
\includegraphics[width=0.95\linewidth]{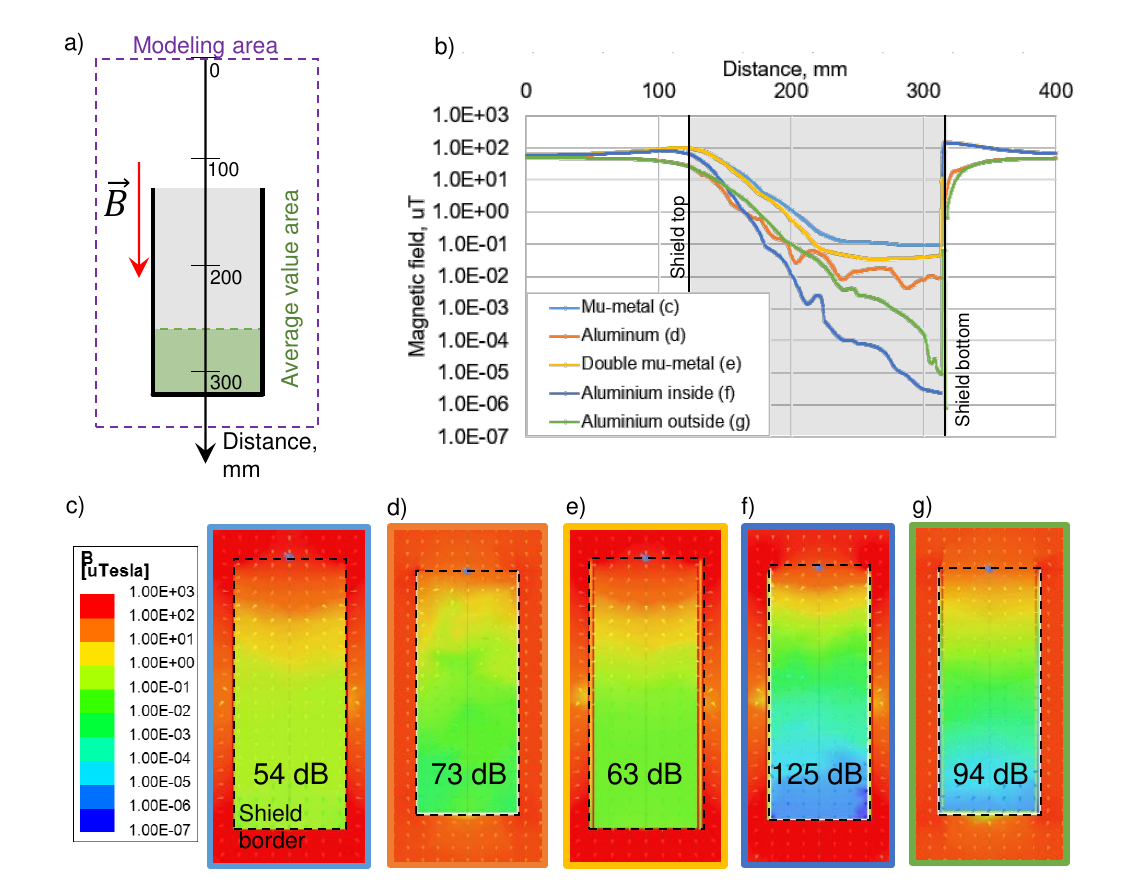}
\caption{\label{fig:wide} Simulation of magnetic field distribution within cylindrical shields: a) calculation model (the external magnetic field of 50 $\mu$T directed along the axis of the cylinders), violet dashed line is modeling area and green area is average value area calculation for magnetic field suppression (MFS) determination; b) comparison of magnetic field strength along the shield axis for various shielding configurations; magnetic field distribution within: c) single $\mu$-metal shield (color bar for magnetic field strength is also valid for d)-g) configurations; the boundaries of the simulated shields are indicated by a black dashed line here and on the d)-g) various configurations; the MFS is indicated in dB on the lower third part of the shielding system here and for d)-g) configurations); d) single aluminum shield, e) double $\mu$-metal shield, f) aluminum shield surrounded with a $\mu$-metal shield, g) $\mu$-metal shield surrounded with an aluminum shield.}
\end{figure*}

\paragraph{The number of layers, the shape and arrangement of the electromagnetic shields.}
Based on the recommendations presented above, it is possible to choose materials for electromagnetic shielding. However, it is also very important to determine the configuration of shielding, i.e. the arrangement and number of shield layers. In order to determine the optimal order of magnetic and superconducting shields for maximum shielding effectiveness, we simulated the magnetic field distribution within cylindrical shields.\\
The magnetic field was simulated using the finite element method. Two cylindrical shields were considered, nested one inside the other. The shields were placed in a vacuum parallelepiped. The external magnetic field of 50 $\mu T$ along the axis of the cylinders was set. Relative magnetic permeabilities were assigned for the $\mu$-metal (70,000), superconductor ($1\cdot10^{-6}$), and vacuum (1). Magnetic induction distribution in the area was simulated. The thickness of the shields is set to 1 $mm$. The heights of the shields are 180 $mm$ and 190 $mm$. Outer diameters are 66 $mm$ and 70 $mm$. In the double-shield configuration, the minimum wall-to-wall separation between shields was maintained at 2 $mm$. The calculation scheme and simulation results are presented in Figure 12. The simulation of magnetic shield configurations yields the magnetic field distribution inside the shielding.\\
For quantitative comparison of different shielding configurations, we calculated magnetic field suppression (MFS) factor within the shielded volume, defined as:

\begin{equation}
MFS=\mid 20lg\frac{B_{average 1/3}}{B_{external}}\mid dB,
\end{equation}
where $B_{external}$ – external magnetic field 50 $\mu T$; $B_{average 1/3}$ – the mean magnetic field value within the lower third section of the shields (where the sample holder is typically positioned).

Modeling results indicate that the optimal configuration for minimizing magnetic field at the SQC position consists of an inner aluminum shield and outer $\mu$-metal shield. When the aluminum shield is positioned inside the $\mu$-metal shield, it operates within a pre-attenuated magnetic field, yielding enhanced overall attenuation. This improvement arises from the complementary shielding mechanisms: $\mu$-metal "pulls over" magnetic fields through its high magnetic permeability, while superconducting materials exclude magnetic fields via the Meissner effect.\\
Placing the superconducting shield externally causes magnetic field lines to distort around the outer shield and concentrate near the opening, potentially compromising sample protection. In contrast, the $\mu$-metal outer configuration provides superior attenuation by absorbing a substantial portion of incident fields before the superconducting shield expels residual flux. An additional consideration is that external placement of superconducting shields may lead to flux trapping phenomena.\\
The combination of magnetic shields should be chosen based on the required external magnetic field suppression level. Regarding target field determination, the required magnetic field suppression should be based on the quantum circuit dimensions. For illustration, consider a frequency-tunable transmon qubit with a SQUID loop area ($S_{SQUID}$) of 30 $\mu m^2$. The magnetic flux quantum of $\Phi_0 = 2.07\cdot10^{-15}Wb$ establishes a reference field:

\begin{equation}
\Phi=B \cdot dS \Rightarrow B_0 = \frac{\Phi_0}{S_{SQUID}} = \frac{2.07\cdot10^{-15}}{30\cdot10^{-12}}\approx 68 \mu T
\end{equation}
This calculation indicates that approximately 69 $\mu$T external field would induce one flux quantum in such a SQUID. Consequently, effective shielding should reduce ambient fields to 0.1-1\% of this value (0.07-0.69 $\mu$T) to ensure residual fields remain negligible and compensable via flux-bias lines. For this particular case, shielding with MFS of at least 60 $dB$ will be required, necessitating either a system with a minimum of two $\mu$-metal shields, or a single superconducting shield.\\
We recommend positioning the shield assembly as close as possible to the sample holder while maintaining minimal shield distances. It reduces the residual magnetic field in the sample space and enables more efficient use of the limited available space in the cryogenic environment.\\
Modeling results demonstrate that open cylindrical shielding systems without lids achieve minimal magnetic field strength in their lower third section. When lid implementation is problematic, optimal shielding performance requires positioning the sample holder as close as possible to the bottom of the shielding system.

\subsection{\label{sec:level2}Additional protection against high-energy radiation}
In certain studies, there have been identified specific shielding techniques for reducing the impact of high-energy radiation, such as cosmic rays. This type of radiation has a high penetrating power, making it difficult to protect SQC from it. In \cite{Vepsalainen2020, Cardani2021}, lead bricks were placed around the dilution refrigerator at room temperature as a part of the shielding system. These bricks do not act as superconducting shields, but rather as a barrier against X-rays and cosmic radiation. Alternatively, in some cases the dilution refrigerator could be located deep underground \cite{Cardani2021}, or in a room carefully shielded against electromagnetic radiation \cite{sullivan2013, Cooper2013}. Although these methods help to additionally reduce the amount of radiation incident on the SQC, they require more sophisticated implementation.

\subsection{\label{sec:level2}Shielding configurations known at present}
\begin{figure*}
\includegraphics[width=0.95\linewidth]{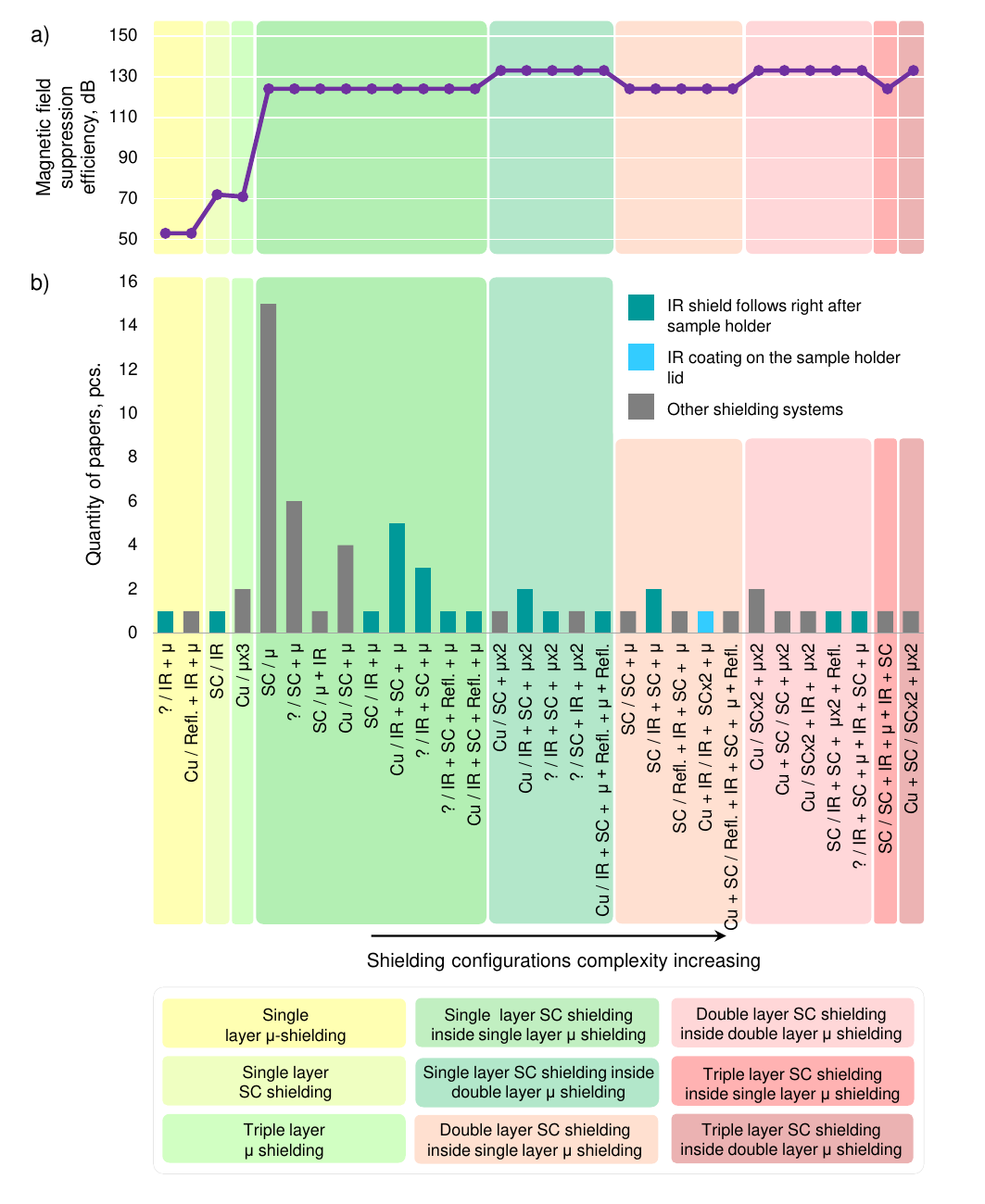}
\caption{\label{fig:wide} A review of existing shielding configurations: a) magnetic field suppression efficiency for shielding systems on the diagram below; b) symbols on the diagram: “Cu” – copper, “?” – not indicated, “/” – separation between the sample holder and shielding system,“SC” – superconducting layer, “IR” – infrared radiation absorbing layer, “$\mu$” – high magnetic permeability layer, and “Refl.” – reflective infrared radiation layer. “Quantity of papers” is the number of referenced papers employing each specific shielding configuration. The color mapping of shielding configurations represents both magnetic field suppression efficiency and system complexity: green indicates configurations achieving high field suppression with relatively simple designs, while red denotes systems that provide comparable suppression but require substantially more complex implementations. Additionally, the column coloring scheme identifies the positioning of IR-absorbing layers within each shielding system.}
\end{figure*}

Let’s consider and analyze the existing shielding systems according to the previously discussed recommendations. Various configurations of multilayered shielding systems are shown in Figure 13. Figure 13 illustrates the hierarchical organization of shielding configurations, presented in sequential order from the sample holder (before the slash) to the innermost shield (immediately after the slash), culminating with the outermost shield in the system. The upper portion of the diagram (Figure 13a) displays the magnetic field suppression efficiency for various shielding systems shown in the lower portion (Figure 13b). These efficiency metrics were derived from our modeling results (Section II.B.3.c).\\
The color coding in the shielding configuration regions represents magnetic field suppression efficiency in combination with system complexity: green denotes configurations with high field suppression and relatively simple designs, while red indicates systems that achieve high suppression but with excessive complexity. Additionally, the column colors specify the positioning of IR-absorbing layers within each shielding system. Together, these visual elements facilitate identification of optimal shielding configurations for SQCs.
We would like to emphasize that the frequency of shielding systems mention in literature does not necessarily correlate with optimal performance. The subsequent analysis in this section aims to justify the prevalence of specific shielding configurations in the literature based on the presented in this paper simulation results.\\
IR shielding is typically implemented using a single absorbing layer\cite{Place2021, Bal2024, Livingston2022} (Figure 14a), with only 10\% of cases employing double-layer configurations \cite{Bengtsson2020, Burnett2019} (Figure 14b). Our modeling results confirm the effectiveness of single-layer shielding positioned immediately adjacent to the sample holder (Figure 6), leading us to highlight these optimal configurations in sea-green in Figure 13b. Computational results further reveals the critical role of the sample holder's interior surface in IR protection for SQCs. While literature examples of this implementation remain limited \cite{Kreikebaum2016} (Figure 14c), we have highlighted these cases through light-blue coding in Figure 13b.\\
\begin{figure*}
\includegraphics[width=0.95\linewidth]{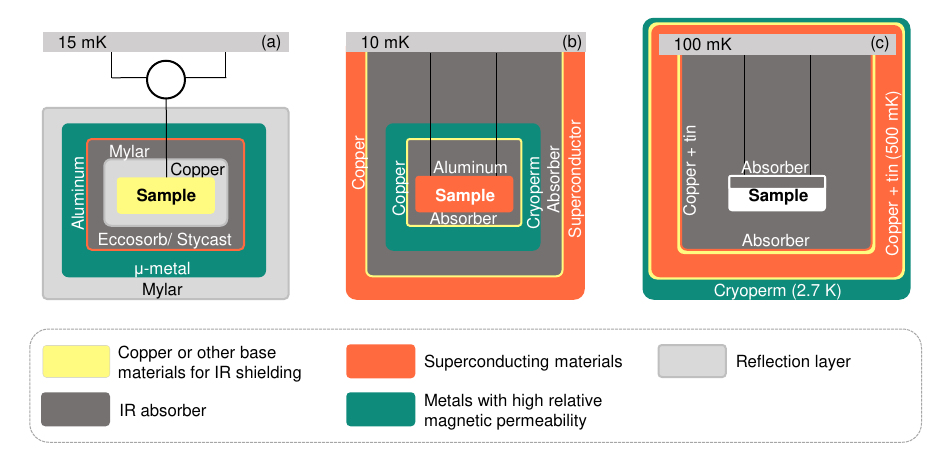}
\caption{\label{fig:wide} Shielding systems with various IR shields: a) single-layer IR shielding combined with superconducting shield\cite{Place2021}; b) a double-layer IR shielding: shields around the sample holder and on the lower cryostat stage\cite{Burnett2019}; c) a double-layer IR shielding: absorbing coating on inner surface of the sample holder and IR shield around it\cite{Kreikebaum2016}.}
\end{figure*}
\begin{figure*}
\includegraphics[width=0.95\linewidth]{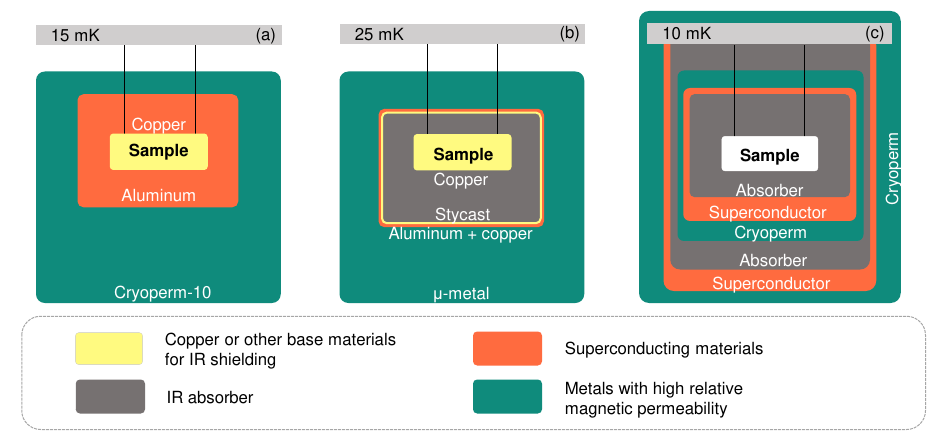}
\caption{\label{fig:wide} Shielding systems with various superconducting shielding configurations: a) a single-layer superconducting shielding following after the sample holder\cite{Yan2016}; b) a single-layer superconducting shielding on copper base and following after IR absorbing layer\cite{Grunhaupt2018}; c) a double-layer superconducting shielding: shields on the second and fifth place from the sample holder\cite{Bengtsson2020}.}
\end{figure*}
\begin{figure*}
\includegraphics[width=0.95\linewidth]{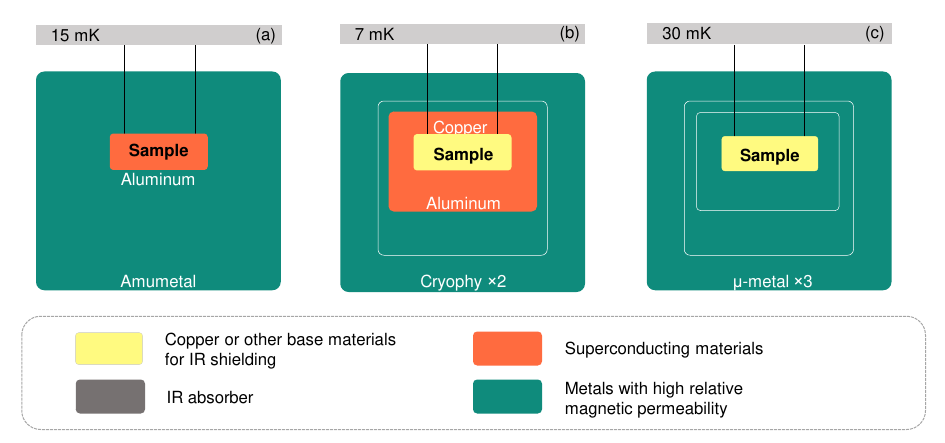}
\caption{\label{fig:wide} Shielding systems with various magnetic shielding configurations: a) a single-layer Amumetal shielding\cite{Reagor2015}; b) a double-layered Cryophy shielding\cite{Asaad2016}; c) a triple-layer $\mu$-metal shielding\cite{Janzen2023}.}
\end{figure*}
Approximately in 80\% of considered works IR shielding is located around the sample holder \cite{Schwartz2016, Gyenis2021, Jin2015, Serniak2019} (Figure 14a). In other cases it surround the lowest stage of the fridge \cite{Burnett2019, Bengtsson2020} as shown in Figure 14b . IR shields are typically positioned immediately after the sample holder \cite{Serniak2019, Dickel2018}. However, in multilayered shielding systems, they can be located at second \cite{Burnett2019} (Figure 14c) or even fourth \cite{Bengtsson2020} shielding layer. Modeling results show that the multilayer IR shielding have the same effect as a single-layer one.\\
Double-layer configurations are implemented in approximately 20\% of superconducting shielding systems\cite{Singh2025, Bengtsson2020improved, Warren2023}. This is well explained by the simulation results, according to which the number of superconducting shielding layers does not affect the shielding quality. These findings are further supported by the modeling results presented in previous sections. Our analysis demonstrates that a single-layer aluminum shield provides adequate protection against external magnetic field fluctuations for SQCs. As evident from Figure 13, approximately in a half of the examined shielding systems utilize superconducting sample holders, which simultaneously function as effective shielding components. This integrated approach represents a significant simplification of the overall shielding architecture.\\
The location of the superconducting shields may vary, most of them directly surround the holder \cite{Jin2015}, while the others located either at lower \cite{Burnett2019, Bengtsson2020} or penultimate stage of the fridge \cite{Kreikebaum2016}. Due to the requirement of the magnetic field to be uniformly distributed at the sample holder place, it is preferable to close the superconducting shield by the lid.\\
Superconducting shields can be positioned either immediately adjacent to the sample holder\cite{Yan2016, Oh2024, Eddins2019} (Figure 15a), or following the IR shielding layer\cite{Grunhaupt2018, Dickel2018, Jin2015} (Figure 15b). In more complex systems comprising four or more shields, superconducting layers may occupy the third or even fifth position relative to the sample holder\cite{Bengtsson2020} (Figure 15c). However, our modeling results showed that multilayered shielding system do not significantly increase shielding efficiency, so a more compact design may be used.\\
Our analysis of the literature reveals that single-layer magnetic shielding \cite{Reagor2015, Braumüller2017, Dodge2023} (Figure 16a) predominates, being employed in 77\% of surveyed studies. Two-layer configurations \cite{Asaad2016, Warren2023, Norris2024, zwanenburg2025} (Figure 16b) were implemented in 16\% of cases, while only a single research group reported utilizing a three-layer shielding system \cite{Janzen2023, Buks2020} (Figure 16c). As it was previously demonstrated, the combination of a superconducting shield inside and a ${\mu}$-metal shield outside provides the most effective protection against magnetic fields.\\
As well as the other types of the shields, magnetic shields are typically placed around the holder \cite{Burnett2019, Kelly2015, Sung2018}. At the same time, there are more options for the location of magnetic shields within the fridge, such as the lower stage \cite{Jin2015}, 2.7 $K$ \cite{Kreikebaum2016} or around the fridge itself \cite{moghaddam2019}.\\
Modeling results demonstrate that effective shielding requires fully enclosed configurations. Therefore, the shield must completely surround the sample holder to achieve optimal performance. In case the magnetic shield is the only one shield used in a system, it typically follows right after the sample holder \cite{Dodge2023, Chapman2023, Oh2024}. In other cases, the magnetic shield may be placed outside of the other shields, such as IR shields \cite{Chou2018, Wu2021} and superconducting shields \cite{Monroe2021, Asaad2016}. In multilayered magnetic shielding systems, ${\mu}$-metal shields can be incorporated between either superconducting or IR shielding layers\cite{Bengtsson2020, Burnett2019}. As it was previously mentioned, for optimal shielding performance, the superconducting shield should be placed inside the ${\mu}$-metal one.\\
Most considered shielding systems meet our requirements for the shielding. Requirements description for the shields based on the modeling results make it possible to identify excessiveness in the shielding system and make it lighter and more compact.

\section{MICROWAVE LINES FILTERING}
As previously mentioned, the optimal protection of SQCs from stray electromagnetic fields can be achieved by the use of shielding in combination with microwave filtering. In the following paragraphs we’ll investigate in more detail the different types of filtering and consider their applications in existing cryogenic setups.\\
All control lines of SQC (here we are focused on SQC based on transmon qubits) can be divided into four categories:
\begin{enumerate}
\item Readout input lines. The typical frequency band for these is from 4 to 8 $GHz$, power required for single qubit readout is around -110 $dBm$ in the case of system without on-chip purcell filters. In multiqubit systems one readout line can be used for readout several (up to 10) qubits, so in this case the power coming on chip is around -100 $dBm$.
\item Readout output lines. The typical frequency band for these is from 4 to 8 $GHz$, typical power coming from chip to the first stage amplifier is around -115 $dBm$ in the case of single qubit readout and around -105 $dBm$ in the case of multiqubit readout. The main feature of readout output line is the necessary of minimizing signal losses from quantum chip to the first stage amplifier for improving readout efficiency, therefore it is undesirable to use attenuators or high-loss IR filter for it.
\item Drive lines (lines for performing arbitrary single qubit rotations). The typical frequency band for these is from 4 to 8 $GHz$, power required for fast single qubit rotations is around -80 $dBm$.
\item Flux lines (lines for in-situ tuning of superconducting qubit frequency). The typical frequency band for these is from DC to 500 $MHz$ (in the case of fast flux-assisted two qubit gates), the values of DC current required to adjust the frequency is on the order of several $mA$.
\end{enumerate}

For high performance of SQC it is necessary to filter the frequencies out of band for all the types of lines listed above. Insufficient filtering of frequencies close to the qubit ones can lead to parasitic single-qubit rotations, degrading the single qubit gate fidelities, as well as undesirable population of possible coupled with qubits parasitic modes, which can lead to additional qubits decoherence. Insufficient filtering of IR frequencies (above 90 $GHz$) can lead to cooper-pairs breaking in superconductor thin film and also degrade the qubit coherence. In this article we focus only on the different filter types used for all lines listed above and made some recommendations for its filtering. The purpose of other microwave components, used in control lines, such as attenuators, circulators, isolators etc., is described in detail in Ref.\cite{Krinner2019}.

\subsection{\label{sec:level2}Requirements for filters}
Filtering is usually used to protect SQCs from stray electromagnetic fields and IR radiation that can propagate in microwave lines. For this purpose, various types of microwave filters covering the SQC operating range (typically, 4-8 $GHz$) are usually used, namely Low-pass filter – LPF\cite{Arute2019, Premkumar2021, Serniak2019, moghaddam2019, Bengtsson2020}, High-pass filter - HPF\cite{Burnett2013, kudra2020} or Band-pass filter - BPF\cite{Buks2020, Lisenfeld2019, Mundhada2020}. This filters are designed to cut off a portion of the signal spectrum outside the denoted passband, suppressing most of the stray electromagnetic fields. For suppressing IR radiation, a special IR filters are usually used. The working principle of such filters based on IR radiation power loss due to eddy currents excitation in filter filler material \cite{WANG2014}.\\
It is necessary to use microwave filters with IR ones, as they cut-off different parts of the signal spectrum. The main requirement for IR filters is as low as possible attenuation in the SQC operating frequency range, smooth transmission characteristic and at the same time high enough attenuation (no less than 20 $dB$) at higher frequencies (above 20 $GHz$).\\
One of the important requirements for microwave components (attenuators, filters, etc.) when operating at millikelvin temperatures is the availability of efficient thermalization. The lack of proper thermalization can lead to a significant increase in the effective temperature of the qubits \cite{yeh2017, Serniak2019}. The issue of thermalization, methods of increasing its efficiency and its influence on qubits effective temperature is considered in detail in \cite{yeh2017, Yeh2019}, therefore, in this review we decided not to focus on this.

\subsection{\label{sec:level2}IR filters}
In works, IR filters may have different designations (IR filter \cite{Chou2018, Opremcak2021, Kelly2015, Schneider2019}, lossy filter \cite{Schwartz2016, Tan2015}, thermalization filter \cite{Kreikebaum2016}, etc.), but all of them are designed for the same purpose: to suppress stray IR radiation in signal lines. The filling material for such filters (see Figure 16) typically consists of an epoxy resin containing fine particles of carbonyl iron (Eccosorb CR-110 \cite{Place2021, Narla2017, Reagor2015}, or, less frequently, CR-124 \cite{Krinner2019}), and metal powder \cite{barends2011} (mainly copper \cite{Colless2018}). Other possible fillers include epoxy resin based on aluminum oxide (Stycast) mixed with bronze powder \cite{deVisser2014}, or without it \cite{Lisenfeld2019}. Specific fillers such as carbon nanotubes may also be used \cite{moghaddam2019}.\\
There are no studies with a direct comparison of the IR filters efficiency with different types of filling material. However, in considered works, Eccosorb CR-110 epoxy is the most widely used in IR filters (see Figure 17).

\begin{figure}
\includegraphics[width=0.95\linewidth]{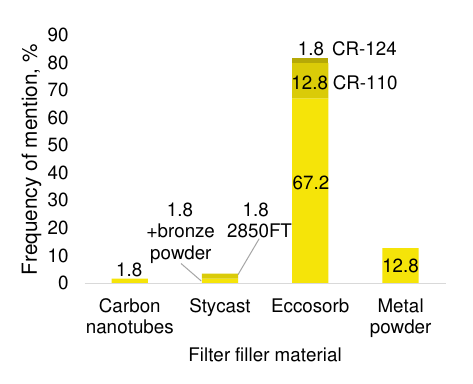}
\caption{\label{fig:epsart} Typical filling materials for IR filters.}
\end{figure}

The attenuation coefficient of Eccosorb CR-110 exhibits a frequency-dependent increase from 5.4 $dB/cm$ at 100 $GHz$ to 90 $dB/cm$ at 1 $THz$\cite{Connolly2024}. Due to its enhanced effectiveness at higher frequencies, even compact Eccosorb filters can effectively suppress Cooper-pair-breaking photons above 100 $GHz$\cite{Connolly2024}.\\
The optimal filter length should be determined both by the required attenuation at the Cooper pair-breaking threshold frequency (e.g., $\sim$100 $GHz$ for 100 $nm$ thick aluminum films), and the maximum permissible attenuation at the qubit control frequency. For instance, in work\cite{Connolly2024} was shown that a 20 $mm$ long filter providing 0.5 $dB$ attenuation at 5 $GHz$ and ensure a charge parity switching rate of 10 $s^{-1}$, corresponding to a relaxation time limit of 100 $ms$.\\
IR filters are employed in all types of signal lines of cryogenic setup. In the following sections we will consider importance of IR filter using in different line types.

\subsection{\label{sec:level2}Drive line filtering (XY control)}
For the SQCs driving, special drive lines are utilized. In some implementations, the SQC driving is carried out through a readout line. However, in this review, we will consider an individual drive lines separated from readout.\\
LPFs or BPFs are essential on qubit drive lines to suppress parasitic harmonics generated during frequency up-conversion. These filters can be placed either at room temperature or at the base of the cryostat. For room-temperature implementation, an isolator must be positioned upstream of the filter to prevent standing wave formation and subsequent control pulse distortion\cite{Li2023}. In contrast, placement at the lowest cryostat stage utilizes attenuators to inherently suppress standing waves\cite{Kelly2015}, rendering additional isolation unnecessary.\\
Additional IR filtering is required for the qubit drive line, as standard microwave LPFs and BPFs provide no specified attenuation outside their operational bandwidth. For instance, the Marki FLP-0750 filter exhibits effective performance only up to 20 $GHz$\cite{Marki}. The IR filter must be positioned between attenuators to ensure proper impedance matching\cite{Krinner2019}. Furthermore, it requires enclosure within a shield anchored to the cryostat's lower stage to prevent infrared radiation leakage from warmer stages into the post-filter segment of the drive line\cite{Serniak2019}.

\subsection{\label{sec:level2}Flux line filtering (Z control)}
The flux line filtering is usually based on application of LPF and IR filters. For this line, as for the drive line, we consider a separated flux line (in some implementations the magnetic flux can be manipulated by a special coil surrounding the sample holder).\\
LPFs are essential on qubit flux lines to suppress noise bandwidth contributions from both warmer cryogenic stages and control electronics. The filter's cutoff frequency must be selected according to the required flux pulse duration - for instance, a minimum 200 $MHz$ cutoff is necessary to support fast two-qubit operations with 15 $ns$ pulses\cite{Foxen2020}. In practice, 500 $MHz$ filters are commonly used\cite{Arute2019} as a trade-off between fast two-qubit operations and the noise bandwidth.\\
IR filters are essential on flux lines to protect against Cooper-pair-breaking radiation. Since LPFs operate via signal reflection (rejecting frequencies above cutoff), this reflection mechanism creates standing waves between the LPF and qubit at operational frequencies, inducing qubit relaxation\cite{Klimov2018}. To suppress these parasitic modes, an IR filter must be inserted between the LPF and qubit\cite{ivanov2023robust}. This IR filter must exhibit minimal reflection at qubit frequencies (less than -15 $dB$). For these applications, Eccosorb CR-124 serves as an optimal IR absorber due to its frequency-selective properties: low absorption (< 0.5 $dB/cm$) below 500 $MHz$ and high absorption ($\sim$ 40 $dB/cm$) at typical qubit frequencies\cite{Ecco124}.
Furthermore, Eccosorb CR-124-based filters enable combined flux and drive line implementations \cite{Li2025, Manenti2021}, as this material simultaneously provides sufficient IR attenuation for flux line protection, and adequate microwave attenuation for drive line.

\subsection{\label{sec:level2}Filtering of readout line input}
Typically, in cryogenic setups one \cite{Córcoles2015, Wu2021, Buks2020, Martin2020, Lachance-Quirion2017} or two filters \cite{Gyenis2021, Schneider2019, Opremcak2021} are used at the input readout line. Less frequently, three filters are installed in a row \cite{Chen2018, Colless2018, Grünhaupt2019}. Microwave and IR filters can be used both individually \cite{Schwartz2016, Yan2016, Grunhaupt2017, Andersen2019} and together \cite{Zhang2018, Lisenfeld2019, Leonard2019}.\\
In most measurement circuits, filters are located outside the shielding \cite{Jin2015, Arute2019, yeh2017, Leonard2018, kudra2020, Wilen2021}. Less frequently, they may be located both inside and outside the shielding \cite{Burnett2019, Premkumar2021, Gao2018, Mundhada2020}. In very rare cases, the filters are only located inside the shielding \cite{Kreikebaum2016}, mounted on the lid of the shield \cite{Hazard2019, barends2011}, or just prior to the sample holder \cite{Cooper2013}. In \cite{Serniak2019} it is indicated that the location of an IR filter inside the shielding can reduce the quasiparticles generation in a qubit. Thus, it is recommended to place the IR filter within a metal shield thermalized on the base stage of the dilution refrigerator. For microwave filters there is no special requirements for positioning in the measurement circuit. Thus, they are typically located outside the shielding. The filtering requirements for readout line input are generally equivalent to those for qubit drive lines, necessitating comparable filter specifications and implementation strategies.\\

\subsection{\label{sec:level2}Filtering of readout line output}
The readout line output requires minimal signal attenuation to preserve the signal-to-noise ratio (SNR), necessitating the use of non-dissipative filters (low-pass or band-pass filters) and isolators. This configuration provides three critical benefits: firstly, suppression of thermal noise propagation from warmer cryostat stages; secondly, prevention of signal reflection from the output line back to the qubit; and thirdly, isolation of parametric amplifier pump leakage.
For IR radiation mitigation, either compact IR filters (to minimize output signal attenuation) or a High-Energy Radiation Drain (HERD)\cite{Rehammar2023} filter should be implemented.

\section{METHODS FOR EVALUATING THE EFFECTIVENESS OF SHIELDING SYSTEMS AND LINES FILTERING}
In order to select a shielding system, which can reliably protect from stray electromagnetic and IR radiation, an assessment of the shielding effectiveness should be done. The main challenge of selecting such system is that the system should be compact having a high shielding efficiency. The shielding effectiveness can be evaluated based on the measurements of specific parameters. The values of these parameters allow the quantitative evaluation of the incident radiation. Several techniques based on the SQC parameters measurement have been proposed in the literature. Based on our review, the following techniques are used: the resonators Q-factor evaluation, the SQC effective temperature measurement, the qubit energy relaxation rate measurement, and the estimation of quasiparticle tunneling rate. Let’s briefly consider all of these methods.

\subsection{\label{sec:level2}Resonator Q-factor ($Q_i$)}
A spontaneous IR radiation with energy exceeding the binding energy of a Cooper pair ($f$ ${\geq}$ 2${\Delta}$/h ${\approx}$ 100 $GHz$ for aluminum thin films \cite{Song2009}) can cause the breakup of a Cooper pair creating the quasiparticles. The resonator Q-factor depends on the density of quasiparticles in superconductor \cite{alexander2025}. Additionally, magnetic vortices may appear in an aluminum thin film under the influence of an external magnetic field, which can also reduce the Q-factor of resonators \cite{Kreikebaum2016, Song2009}. Therefore, the Q-factor of a resonator can be an indicator of how much of spontaneous IR radiation and external magnetic fields reach the sample.\\
In \cite{Kreikebaum2016}, different shielding systems were compared by the values of measured resonators Q-factor. The results showed that the addition of a cryoperm shield at the 2.7 $K$ stage of the dilution refrigerator can increase the resonator internal Q-factor by a factor of two. However, to demonstrate this effect, the researchers created a magnetic field of about 200 ${\mu}T$ inside the dilution refrigerator. The authors did not observe a significant improvement of the resonators internal Q-factor, while increasing the complexity of the shielding.\\
In \cite{barends2011}, it was found that ensuring the light-tightness of the shielding system can be achieved by using an indium gasket between the shield and its lid, which lead to an increase in the Q-factor of a coplanar waveguide resonators from $10^5$ to $10^6$. An important detail of this experiment was that the sample holder was surrounded by a shield with a relatively high temperature of 4 $K$. It is also should be noted, in some experiments in \cite{barends2011}, a special heat source directly radiated the sample was used, which intentionally created a more challenging conditions for the resonators’ performance.\\
Generally, the resonator Q-factor is not the best indicator for evaluating shielding effectiveness, because it requires an external magnetic field or IR radiation source, indicating that it has a low sensitivity.

\subsection{\label{sec:level2}Effective temperature of the quantum system ($T_{eff}$)}
In several studies \cite{Martinis2009, Aumentado2004, Shaw2008, Vool2014}, it has been noted that the measured excited state population (or effective temperature, $T_{eff}$) of a quantum system is significantly higher than the theoretical one. Such unexpected increase in the measured effective temperature is assumed to be due to the presence of nonequilibrium quasiparticles. The quasiparticles can be induced by a stray microwave noise, cosmic rays, or an IR radiation from warmer stages of a dilution refrigerator \cite{Wenner2013}.Therefore, the effective temperature of a qubit can be used as a metric for evaluating the effectiveness of the shielding from spontaneous IR radiation and stray electromagnetic fields. In Ref. \cite{Kitzman_2022}, the effectiveness of an IR shielding with two different absorbing coatings was evaluated based on the qubit effective temperature.\\
This method may be useful for determining the best IR absorber for the shielding, but it is probably not suitable for comparing different configurations of magnetic shielding.

\subsection{\label{sec:level2}Energy relaxation rate of the qubit (${\Gamma}_1$)}
It is known that the energy relaxation rate of a quantum system is composed of two components: the relaxation rate due to the quasiparticles, and the relaxation rate due to a combination of other factors such as radiation, cosmic rays, dielectric losses, and the two-level systems fluctuations \cite{Vepsalainen2020}. Having determined the contribution of each factor to the relaxation rate, it is possible to use this rate as a sensor for the normalized quasiparticle density. As mentioned earlier, the quantity of the quasiparticles is directly correlated with the amount of radiation. Using a calibrated radiation source, the contribution of other sources (cosmic rays, dielectric losses and two-level systems fluctuations) to the energy relaxation rate has been determined in \cite{Vepsalainen2020}. As a result of these measurements, it has been shown that protection from cosmic rays presents a challenging task. Specialized shielding systems designed for protection from cosmic rays do not significantly reduce the energy relaxation rate of the SQC. The authors in \cite{Vepsalainen2020} suggest, that the placement of a dilution refrigerator in a deep underground location and in a room well protected from electromagnetic radiation may be able to accomplish this task.\\
This method is applicable only if there is a calibrated radiation source. Otherwise, it will be challenging to separate the contributions from quasi-equilibrium quasiparticles and other mechanisms, for example, strongly coupled TLS.

\subsection{\label{sec:level2}Quasiparticles tunneling rate (${\Gamma}_{QP}$)}
As far as the quasiparticles generation is associated with the incident radiation \cite{Martinis2009, Catelani2011, Catelani2012}, the value of the quasiparticles tunneling rate through a Josephson junction can be an indicator of how much radiation penetrates into the space surrounding the qubit. The main parameter used in this method is the charge parity lifetime ($T_P$), which can then be converted into a quasiparticle tunneling rate. Measuring the quasiparticle tunneling rate can be done by mapping the charge parity ($P$) onto the quantum system state \cite{Ristè2013}, or directly from the dispersive shift of the resonator \cite{alexander2025}.\\
Using this method, several experiments were done \cite{gordon2022environmental, Connolly2024, Serniak2019}. 
Authors of \cite{gordon2022environmental} have carried out research of various shielding systems and their comparison. As a result, it was demonstrated that different shielding systems may not exhibit significant variations in qubit lifetimes of up to 400 ${\mu}s$ in the presence of the mixing chamber shield, but they do influence the tunneling rate of quasiparticles. Thus, the tunneling rate of quasiparticles is a more sensitive parameter than qubit relaxation rate for shielding efficiency evaluation. In \cite{Connolly2024} it was tested how the length of the IR filter affect the quasiparticles tunneling rate. It was found that the first 0.9 $cm$ of the IR filter is the most effective for quasiparticle tunneling rate decreasing. In \cite{Serniak2019} it was shown that considered method is applicable for definition of IR filter position in cryogenic setup. Author of \cite{Serniak2019} approve that IR filter is better to mount as closer as possible to SQC.\\
All of these results demonstrate that quasiparticle tunneling rate can be used as universal parameter for effectiveness evaluation of both shielding system and filtering of microwave lines. We recommend using the charge parity measurement method as the most accurate one.

\subsection{\label{sec:level2}Other methods of shielding system effectiveness evaluation}
An alternative way for shielding effectiveness estimation can be done by using detectors designated for each type of parasitic factor, e.g., IR radiation and stray electromagnetic fields. Such an approach could allow us to estimate the effectiveness of the shielding constituent parts. Unfortunately, it is not always possible to fabricate a special device and set up a measurement procedure. Moreover, IR detectors are typically fabricated not from the same materials as SQC, that cause additional technological disbursements. Another way for shielding testing is using SQUIDs for electromagnetic noise detection. It can be a SQUID magnetometer \cite{Kawai_2017, Schmelz2016} or the qubits themselves \cite{Braumuller2020}. However, it is out of the scope of this paper.\\

\section{SUMMARY AND OUTLOOK}
In this review, we have described the requirements for the shielding system to protect the SQC from external IR radiation and stray electromagnetic fields. We have considered a variety of existing shielding systems and compared them to our requirements. We have also considered different techniques for shielding system effectiveness evaluation.

\subsection{\label{sec:level2}Recommendations for shielding and filtering}
The conclusions from the previous paragraphs are that there are no specific recommendations for shielding superconducting quantum circuits. Nevertheless, such recommendations can be made using theories of electromagnetic shielding and heat transfer by radiation. Based on these theories and the modelling results, it is possible to identify the most effective shielding system. As a result, the following recommendations were formulated:

\begin{enumerate}
    \item The shielding system should comprise a minimum of three layers: an IR shield, a superconducting shield, and a ${\mu}$-metal shield. However, it is important to note that simpler shield configurations generally provide better cooling efficiency for the entire system.
    \item IR shields must be light-tight to prevent the penetration of infrared radiation. The same requirement applies to the sample holder, which serves as the first stage of IR radiation protection. For superconducting and ${\mu}$-metal shields, a closed design is recommended to ensure magnetic field uniformity throughout the shield volume. If this is not feasible, the sample holder should ideally be positioned in the lower third section of the shields.
    \item In order to create a more compact shield, shielding functions may be combined. Specifically, an IR shield can be combined with a superconducting one by depositing a superconducting layer on a copper base.
    \item To achieve high-efficiency IR shielding, the shield should be constructed with a high-thermal-conductivity base material, such as copper or superconductor-coated copper. For SQC shielding, the sample holder itself may suffice either with an absorbing inner coating (for floating qubits) or without (for other cases). An additional IR shield around the sample holder is advisable if passive measurement circuit elements are nearby. This shield should be positioned as close to the SQC as possible. For the IR-absorbing inner coating of the shield or sample holder, a mixture of Stycast epoxy and $\sim$1 $mm$ SiC particles is recommended.
    \item Since the sample holder acts as the first shielding layer, its fabrication should follow the same guidelines as those for IR shields (see Recommendations 2-4). The sample holder’s design must be adapted to the SQC configuration and the cryogenic setup’s wiring requirements.
    \item For optimal magnetic field protection, shielding configurations should be selected based on the required field suppression level for the specific SQC architecture. Regarding single-layer shielding, superconducting shields maintain lower internal magnetic fields compared to $\mu$-metal shields. For double-layer systems, the optimal configuration combines an inner superconducting shield with an outer $\mu$-metal shield, achieving maximum field suppression.
    \item Superconducting shields can be implemented in two ways. The first method involves a single-layer superconducting aluminum shield following after IR shield. The second method utilizes a combined shield consisting of a superconducting layer on a copper light-tight shield coated by an absorbing material and surrounding the sample holder.
    \item The ${\mu}$-metal shield should be made of a material with a high relative magnetic permeability (100,000 or more), placed right after the IR shield and the superconducting one.
    \item To achieve uniform magnetic field distribution within the shielding system, a fully enclosed configuration (including a lid) is recommended. When this is not feasible, the sample holder containing the SQC should be positioned in the lower third section of the shielding volume, where magnetic field is minimal.
    \item Magnetic components of the cryogenic setup (e.g., isolators, circulators) should be positioned outside the shielding system. Special design features, such as waveguide integration on the shield lid, can further mitigate magnetic field effects on the quantum circuit.
    \item If cosmic rays have an undesirable effect on SQC performance, it may be reasonable to place the dilution refrigerator inside a shielded laboratory or in a deep underground location.
    \item The best protection of the superconducting quantum circuit from stray electromagnetic fields and IR radiation can be achieved by combining the shielding and microwave filters. Effective filtering of all qubit microwave lines (drive, flux, and readout) requires both microwave and IR filters. For drive lines and readout inputs, either "LPF+IR" or "BPF+IR" filter combinations are recommended, where the LPF or BPF should be selected based on the qubit operating frequency. Eccosorb CR-110 epoxy serves as the optimal IR filter filler for these applications.\\
    For flux lines, an "LPF+IR" configuration is preferred. The LPF cutoff frequency must be optimized according to the target flux pulse duration, with Eccosorb CR-124 providing superior performance as the absorber for IR filters in this case.\\
    The readout output line requires similar filtering to the drive and readout input lines ("LPF+IR" or "BPF+IR"). LPF/BPF filters should be matched to the qubit operating frequency, and the IR filter should be minimized in size to reduce output signal attenuation on condition of adequate IR suppression.
\end{enumerate}

All the recommendations given enable the creation of a high efficient and compact shielding system that protects the SQC from stray electromagnetic fields and IR radiation.

\subsection{\label{sec:level2}Method for shielding testing}
The shielding system effectiveness can be most accurately evaluated by the method based on the measurement of the quasiparticle tunneling rate. This method provides a higher level of resolution compared to the other methods and allows for choosing different shielding and filtering systems.

\section{REFERENCES}

\nocite{*}
\bibliography{references}

\end{document}